\documentclass[review]{elsarticle}

\usepackage{cancel}
\usepackage{amsmath}
\usepackage{multirow}
\usepackage{booktabs} 
\usepackage{listings}

\usepackage[ruled]{algorithm2e} 

\usepackage{hyperref}

\journal{Applied Soft Computing}

\begin{document}

\begin{frontmatter}

\title{Evolutionary Design of the Memory Subsystem\footnote{This research has been partially supported by the Ministerio de Econom\'ia y Competitividad of Spain (Grant Refs. TIN2015-65460-C2 and TIN2014-54806-R) and also by the European Regional Development Fund (FEDER) under project EphemeCH (TIN2014-56494-C4-{1,2,3}-P), and Junta de Extremadura FEDER, project GR15068.}}

\author{Josefa~D\'{i}az~\'{A}lvarez}
\address{Dpt. of Computer Architecture and Communications, University of Extremadura, Spain}

\author{Jos\'{e} L. Risco-Mart\'{i}n}
\address{Dpt. of Computer Architecture and Automation, Complutense University of Madrid, Spain}

\author{J. Manuel Colmenar}
\address{Dpt. of Computer Science, King Juan Carlos University, Spain}

\begin{abstract}
The memory hierarchy has a high impact on the performance and power consumption in the system. Moreover, current embedded systems, included in mobile devices, are specifically designed to run multimedia applications, which are memory intensive. This increases the pressure on the memory subsystem and affects the performance and energy consumption. In this regard, the thermal problems, performance degradation and high energy consumption, can cause irreversible damage to the devices.

We address the optimization of the whole memory subsystem with three approaches integrated as a single methodology. Firstly, the thermal impact of register file is analyzed and optimized. Secondly, the cache memory is addressed by optimizing cache configuration according to running applications and improving both performance and power consumption. Finally, we simplify the design and evaluation process of general-purpose and customized dynamic memory manager, in the main memory. To this aim, we apply different evolutionary algorithms in combination with memory simulators and profiling tools. This way, we are able to evaluate the quality of each candidate solution and take advantage of the exploration of solutions given by the optimization algorithm.We also provide an experimental experience where our proposal is assessed using well-known benchmark applications.
\end{abstract}


\end{frontmatter}

\section{Introduction}
Memory hierarchy has a significant impact on performance and energy consumption in the system. This impact is estimated about $50\%$ of the total energy consumption in the chip~\cite{Kandemir2006}. This places the memory subsystem as one of the most important sources to improve both performance and energy consumption. Concerns such as thermal issues or high energy consumption can cause a significant performance degradation, as well as irreversible damages to the devices therefore increasing the energy cost. Previous works have shown that saving energy in the memory subsystem can effectively control transistors aging effect and can significantly extend lifetime of the internal structures \cite{Mahmood2014}. 

Technological changes combined with the development of communications have led to the great expansion of mobile devices such as smartphones, tablets, etc. Mobile devices have evolved rapidly to adapt to the new requirements, giving support to multimedia applications. These devices are supplied with embedded systems, which are mainly battery-powered and usually have less computing resources than desktop systems.

Additionally, multimedia applications are usually memory intensive, so they have high performance requirements which implies a high energy consumption. These features increase the pressure on the whole memory subsystem. 

Processor registers, smaller in size, work at the same speed than the processor and consume less energy compared with other levels of the memory subsystem. However, the energy consumption and access time rise when the file size increases due to a higher number of registers and ports. 

Regarding the cache memory, it has been identified as a cold area in the chip, although the peripheral circuits and the size of the cache are the most influencing factors to cause a temperature increase~\cite{Meterelliyoz-2010}, facing the accesses to the cache memory because of specific applications. However, cache memory affects both performance and energy consumption. In fact, energy consumption of the on chip cache memory is considered to be responsible of $20\%$ to $30\%$ of the total consumption in the chip~\cite{Varma05}. A suitable cache configuration will improve both metrics. 

In terms of performance, the main memory is the slowest component compared with the cache memory and processor registers. Running programs request the allocation and deallocation of memory blocks, and the Dynamic Memory Manager (DMM) is in charge of this task. Current multimedia applications have highly dynamic memory requirements, so optimizing the memory allocator is a crucial task. Solving a memory allocation request is a complex task and the allocation algorithm must minimize internal and external fragmentation problems. Therefore, efficient tools must be provided to DMM designers for evaluating the cost and the efficiency of DMMs, facilitating the design of customized DMMs.

\begin{figure*}[ht!]
  \centering
  \includegraphics[width=0.9\textwidth]{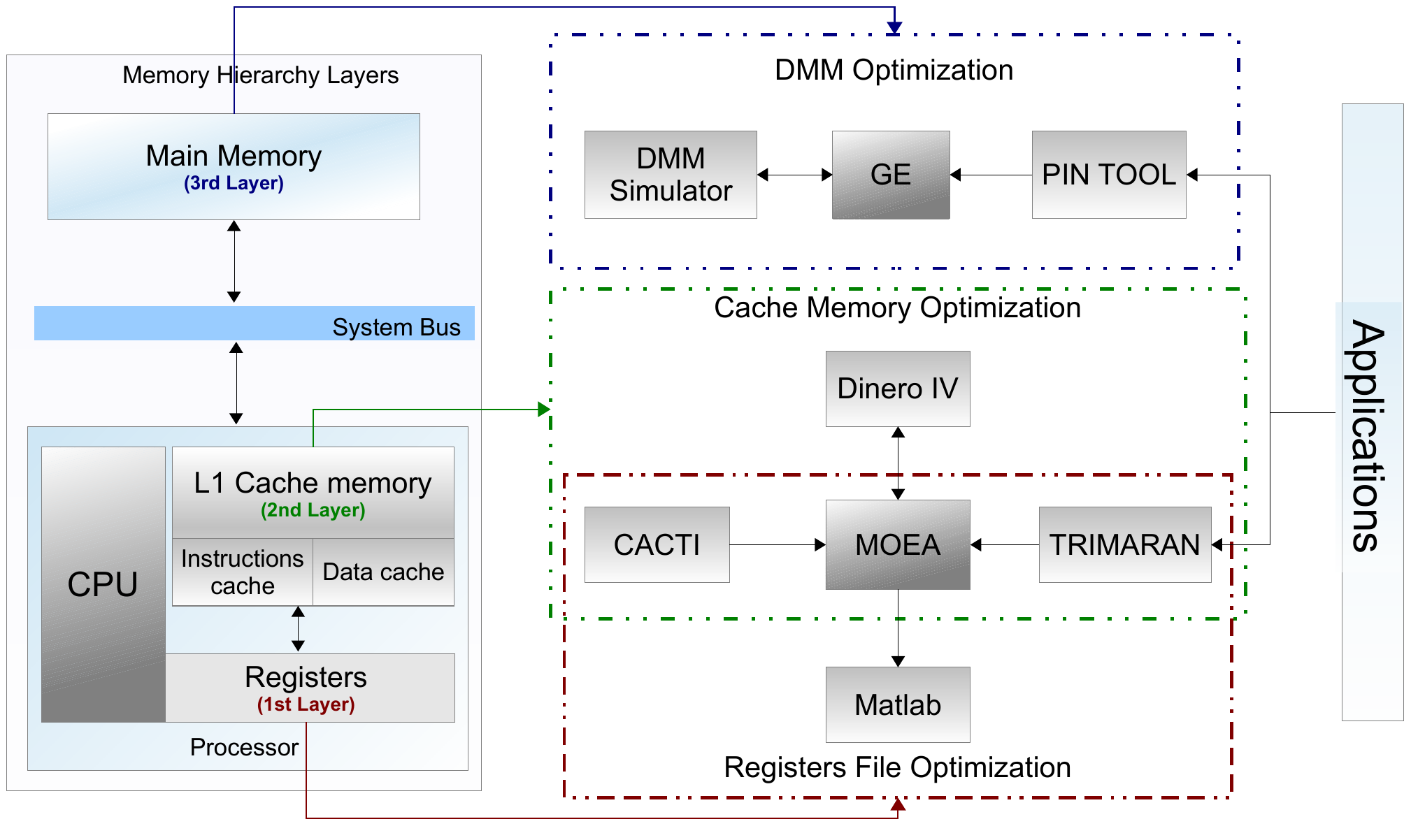}
  \caption{Memory subsystem layers and tools involved in this optimization methodology. First layer is the register file structure, the second one corresponds to the cache memory and 
  the third layer is the dynamic memory which works on the main memory. 
  }
  \label{figure:method_layer} 
\end{figure*}

In this paper we present a methodology based on Evolutionary Algorithms (EA), which is divided into three layers tackling different components of the memory hierarchy and performing the optimization process of each layer according to the running applications. Then, the first layer is the registers file, the second is the cache memory and the last one is the DMM, which works on the main memory. \figurename{~\ref{figure:method_layer}} shows the three optimization layers surrounded with different dashed lines, and the tools involved within each optimization process, which will be deeply explained in the rest of the paper.

In a previous work~\cite{DiazAlvarez-2015}, we presented an approach based on Grammatical Evolution (GE) with a wide design space, where the complete set of parameters defined is considered and a specific cache memory configuration was chosen as a baseline. The GE approach had good results, in the absence of other results to be compared with. The problem is clearly multi-objective and thus the GE approach considered a weighted objective function.  Hence, the optimization problem was later addressed through a multi-objective approach with NSGA-II~\cite{DiazAlvarez2016}. On the one hand, this approach was customized with a fixed cache size for both the instructions and data cache.  On the other hand, a different cache memory configuration was used as the baseline. Thus, GE and NSGA-II approaches use a different set of parameters. As a consequence, results could not be directly compared in order to take a decision. 

In this paper we provide several new contributions regarding the cache design. Firstly, we perform the experiments using the NSGA-II algorithm in the same conditions of the GE proposal, both the design space and the baseline. This configuration allows a direct comparison among both algorithms. Additionally, two baseline caches, included in general purpose devices, are added to the analysis because the first one belonged to a specific purpose device. Finally, we have added a statistical test to verify the relevance of the results. Therefore, this work completes the set of tests previously made and provide us enough information to decide the algorithm to be applied in the cache design optimization. 

In addition to the cache design, we propose in this paper to apply evolutionary techniques to the register file configuration and the DMM which, considered in conjunction with the cache, comprise the whole memory subsystem in a computer. For both the register file and the DMM we propose the algorithms, perform the experiments and analyze the results on both objectives of our fitness function: execution time and energy consumption. Besides, we have incorporated statistical tests to verify the relevance or our results in both the register file and the DMM optimizations. Up to our knowledge, a complete 3-layer approach as the one we propose has not been reported previously in the literature.

We have also focused our experiments on the ARM architecture, which is present in many of the current embedded multimedia systems. Selected applications have been adapted in order to better fit to each one of the memory layers that we optimize. As we will show later in this work, the cache memory policies and the DMMs are most sensitive to improvement.

All the algorithms are coded in Java using the JECO library \cite{ABSysJECO}. Besides, the experimentation has been conducted in a computer provided with an Intel i5 660 processor running at 3.3 GHz, with 8GB of RAM and using the Ubuntu Desktop 14.04 operating system. 

The rest of the paper is organized as follows. Next section summarizes the related work on the topic. Section~\ref{sec:TEP_model} describes the thermal, performance and energy models applied. Section~\ref{sec:registers} addresses the thermal impact on the processor registers. Section~\ref{sec:cache} presents the optimization process aimed to automatically design cache configurations in order to improve performance and reduce energy consumption. Section~\ref{sec:main_memory} describes the optimization process to automatically 
evaluate and design customized DMMs, which will improve performance and reduce the memory fragmentation problem. In Section~\ref{sec:Conclusions}, we present our conclusions and describe the future work. 

\section{Related work}

Many works can be found in the literature regarding the memory optimization. Next, we will review the closest literature to our work, separating the papers into the three memory layers we have studied.

Concerns about thermal problems, performance degradation and high energy consumption are neither new nor insignificant in the memory subsystem. 
The register file is identified as a component that consumes high energy, between $15\%$-$36\%$ in embedded processors~\cite{Tabkhi-2014}. Multimedia applications increase the exchange of information between the register file and the next level of the memory hierarchy. As mentioned in~\cite{Abella-2003}, the number of concurrent accesses is increased and thereby the chip temperature and the need for power dissipation. So, a lower energy consumption reduces the temperature and the need of power dissipation. Thus, system reliability and performance are improved. This problem has been addressed with different hardware and software techniques.

Atienza et al.~\cite{Atienza-2006} are focused on DSP (Digital Signal Processing) and ASIP (Application-Specific Instruction-Set Processor) architectures, specially led to multimedia applications in embedded systems. The authors apply the DVS (Dynamic Voltage Scale) technique to change to low-power state the unused registers. The energy consumption improvement they report is over $60\%$. This work focuses on the energy consumption, but not on the temperature. Zhou et al. ~\cite{Zhou-2009} assign to the compiler the task of distributing the registers access, within the limits of each registers file, in a multi-bank organization. This process is made after the traditional allocation phase and during the registers allocation phase. The proposal, designed  for a limited set of VLIW or RISC architectures, reduces the power density and the peak temperature between $4^{\circ}$C and $7^{\circ}$C. Recently, Sabry et al.~\cite{Sabry-2010} proposed a new mechanism to distribute uniformly the register accesses. This approach, implemented in a commercial compiler, reduces hot spots by $91\%$ and the mean and peak temperature by $11\%$. 

In contrast with previously mentioned works, we use an analytic process to measure the thermal impact of register accesses on the processor temperature. After that, we apply an evolutionary optimization algorithm, which generates a re-assignation that exchanges register accesses to the register file. Therefore, highly accessed registers are spaced far apart, and as a consequence, temperature is reduced.

Cache memory behavior is determined by its parameters, which form the so-called cache configuration. Therefore, the problem is to find the optimal cache configuration for a set of applications running on a system. This will not only improve the performance and the energy consumption, but will also provide long-term reliability. 

Cache memory optimization has been widely addressed. New developments in well-known techniques allow optimizing the cache memory. Wang et al.~\cite{Wang-2014} presented Futility Scaling, a new replacement-based partitioning scheme. This scheme controls the size of the partition and it is able to maintain both large partition and high associativity.

Adegbija and Gordon~\cite{Gordon-2014} designed a phase-based cache tuning algorithm for multimedia applications to determine the best cache configuration for each execution phase.  Phase classification breaks applications execution using a fixed tuning interval. The proposed algorithm analyses each configuration for one interval to determine the best configuration or the next one to be explored. Wang et al.~\cite{Wang-2012} proposed dynamic cache reconfiguration for real time embedded systems. They minimize energy consumption performing a static analysis at runtime. Zang et al.~\cite{Zang2013} applied \textit{way-concatenation} to reconfigure cache in embedded systems by software and minimize the energy consumption. 


Recently, new technologies and core-based processor technologies, such as ARM946E-S TM~\cite{ARMreconf2014}, allow changing the cache configuration for each application. Changes affect the main parameters: capacity, block size and associativity. However, every application has a different memory access pattern. Hence, an efficient algorithm is needed to determine the optimal values for each parameter on each application. 

Previously mentioned approaches optimize a few number of parameters: cache size, block size and associativity, each one with a few possible values. On the other hand, dynamic reconfiguration adds complexity to the memory subsystem design, usually being penalized with extra cost in execution time. Besides, concurrent applications increase this penalty because of the multiple calls to the reconfiguration.


In relation to static profiling, Feng et al.~\cite{Feng2011} applied  a new cache replacement policy to perform the replacement decision based on the reuse of information of the cache lines and the requested data developing to reuse information predictors: a profile-based static predictor and a runtime predictor. However, these approaches only improve the replacement algorithm of the cache.
Recently, Gordon-Ross et al. \cite{Gordon-Ross-2012} studied the interaction between code reordering and cache configuration, obtaining excellent results. However, this technique is applied to the instructions cache, and our systematic optimization method is applied to the full configuration of both the instructions and data caches.



None of these approaches simultaneously optimized cache performance and energy consumption for a target set of applications, as our methodology does. Our proposal optimizes cache size, line size, associativity, replacement algorithm and search algorithm for both instructions and data cache, and also write policy for data cache.

Optimizing the dynamic memory management subsystem is considered a crucial task to efficiently execute multimedia applications on all kind of systems, including embedded 
systems. In this regard, Del Rosso~\cite{DelRosso-2006} evaluated the performance of different DMMs on embedded real time systems. Metrics applied are the internal 
fragmentation and a new metric named \textit{performance speed metric}. However, energy consumption is not analyzed. Atienza et al. in~\cite{Atienza-DMM-2006a} proposed a 
method to evaluate the memory use and energy consumption by a DMM, but it needs to be implemented and integrated in the target application. Risco et al.~\cite{RiscoMartin2009b} 
presented an optimization algorithm, based on Grammatical Evolution (GE), to design customized DMMs. Each DMM was evaluated by a DMM simulator~\cite{Risco-2011}. Although 
this approach allows us to improve the average performance, memory use and energy consumption of the memory subsystem, the classification process returns a complex taxonomy of 
DMMs. Moreover, the applications profiling is made by overloading \textit{malloc()} and \textit{free()} functions, which requires applications to be modified and re-compiled 
for each target application, which is a time consuming and error prone task. 

We present a methodology, which does not need to be integrated in the target application, to automatically evaluate and design customized DMMs. Our methodology is based on GE and also performs a static profiling of applications. In fact, our optimization process produces customized DMMs that are better or equal than well known general purpose DMMs, such as  Kingsley (used in Windows systems) and Lea (used in GNU/Linux systems). The first one is considered the fastest, and the second one, the more efficient with respect to the memory usage.

As seen, we propose the optimization of the complete memory subsystem under both the execution time and energy consumption objective functions. After a thorough review of the literature, we have not found any similar approach to compare with. Hence, as we will show in the experimental experience, we have compared our results with baseline configurations coming from the state of the art of the memory design.

\section{Thermal, energy and performance model}
\label{sec:TEP_model}
The proposed framework is based on the simulation of performance and energy consumption models. These models are used as the input for the optimization algorithms, which find an optimized design. In order to address these works, we have to apply thermal, energy and performance models, which are next described. 

\subsection{Thermal model}
\label{sec:thermal_model}

Estimating the thermal impact in an Integrated Circuit (IC) needs the simulation of thermal conduction between power sources (transistors and interconnects) and heat sinks to the ambient environment. This is analogous to modeling electrical conduction and it is governed by the known Fourier's law. Taking into account one dimension, the thermal problem can be addressed through differential equations~\cite{Yang-2007}. According to Brooks et al.\cite{Brooks-2007} the following equation governs thermal conduction in a chip: 

 \begin{equation}
 \label{eq:dis_calor} 
  \rho c \frac{\partial T(\vec r,t)}{\partial t} = \nabla \cdot \left[K \left(\vec r\right)\nabla T \left(\vec r.t\right) + p\left(\vec r,t\right)\right]
 \end{equation}
 
$\rho$ is the material density, $c$ represents the mass heat capacity, $T \left(\vec r,t\right) $ and $K \left(\vec r\right)$ are the temperature and thermal conductivity of the material at position $\vec r$ and time t, and $ p\left(\vec r,t\right)$ is the power density of the heat source. Extended information is available in the original reference. 

This analysis is carried out applying the method of finite differences. Thus, elements are discretised by dividing the IC area into single elements of equal size. The thermal component of each element can be individually calculated depending on the time, material, power dissipation and temperature of its neighbors. Thus, every element interacts with the rest through heat dissipation, and it has a power dissipation, temperature, capacitance and thermal resistivity to adjacent elements. Thus, the thermal impact in an internal point of the chip can be solved by using~\eqref{brooks_2007}.

\begin{equation}
\label{brooks_2007}
 \frac{CdT(t)}{d_t}+ AT(t) = Pu(t)
\end{equation}

where C is the thermal capacitance matrix as an $N x N$ diagonal matrix, $A$ represents the thermal conductivity matrix as a sparse matrix sizing $N x N$. $T(t)$ and $P$ are $N x 1$ temperature and power vectors and $u(t)$ is the time step function.  We apply steady-state thermal analysis, that means the heat flow and power consumption do not vary over time. Hence, terms that depend on time disappear, and Equation~\eqref{brooks_2007} becomes Equation~\eqref{eq:cal_temp}, where the IC temperature, represented with a set of cells, is estimated based on the individual register power and its thermal conductivity.

\begin{equation}
\centering
\label{eq:cal_temp}
 A T = P \rightarrow T = A^{-1}P
\end{equation}

 $T$ represents the temperature calculated as $A^{-1}$, which is the inverse of $A$, multiplied by the power vector $P$.

\subsection{Energy model}
In order to address the register file and cache memory optimization, we need to estimate the energy consumption because of the number of accesses to the register file and cache memory structures during the execution of a given set of multimedia applications.

We have to determine the energy consumed per access to the register file structure to estimate the energy consumption of this structure. In this context, we apply the model 
detailed in~\cite{Qa2009}, which describes how to measure cache energy consumption and performance based on a limited number of cache accesses. 
Authors defined this model as simple, and suitable to measure energy and performance improvement for reconfigure non-cache systems. Energy model is explained according to 
Equation~\eqref{eq:general_energy}, where terms directly related to the cache memory can be drop, given that this structure is not addressed in the register file characterization.

 \begin{small}
 \begin{equation}
 \label{eq:general_energy}
  E_{total}=E_{read} + E_{write} + \cancel{E_{leak(std)} + E_{c->m} +E_{mp} + E_{misc}}
 \end{equation}
 \end{small}

 Thus $E_{total}$ is the total energy consumption in Joules (J), $E_{read}$ and $E_{write}$ correspond to the energy consumption by read and write register file accesses, which are computed by equations~\eqref{eq:e_read} and~\eqref{eq:e_write}. In those equations, $n_{read}$ is the number of read accesses, $E_{dyn\_read}$ is the dynamic read energy, $n_{write}$ represents the number of write accesses and $E_{dyn\_write}$ is the dynamic write energy. $n_{read}$ and $n_{write}$ are computed during the profiling phase of the application. 
 
\begin{eqnarray}
E_{read} = &n_{read} * E_{dyn\_read} \label{eq:e_read}\\
E_{write} = &n_{write} * E_{dyn\_write} \label{eq:e_write}
\end{eqnarray}

In order to address cache memory optimization, we apply the energy and performance model described in~\cite{Janapsatya-2006}. For the sake of clarity, next section briefly explains the performance model. Interested readers can find detailed information in the original reference. Energy model is described by Equation~\eqref{eq:energy1}, which determines the energy consumption for a cache configuration.

\begin{scriptsize} 
\begin{eqnarray}
Energy&=\cancel{execTime \times CPU_{power}} + I_{access} \times I_{access\_energy} + D_{access} \times D_{access\_energy} + \nonumber \\
& I_{miss} \times I_{access\_energy} \times I_{line\_size} + D_{miss} \times D_{access\_energy} \times D_{line\_size} + \nonumber \\
 & I_{miss} \times DRAM_{access\_power} \times  (\small{DRAM_{access\_time}} + \small{I_{line\_size}} \times \frac{1}{\small{DRAM_{bw}}}) + \nonumber \\
 &D_{miss} \times DRAM_{access\_power} \times (\small{DRAM_{access\_time}} + \small{D_{line\_size}} \times \frac{1}{\small{DRAM_{bw}}}) \label{eq:energy1}
 \end{eqnarray}
 \end{scriptsize}
 
where $DRAM_{access\_power}$ is the power consumption for each DRAM access, and $I_{access\_energy}$  and $D_{access\_energy}$ correspond to the energy consumption for 
instructions and data cache accesses, respectively. Terms $I_{access} \times I_{access\_energy}$ and $D_{access} \times D_{access\_energy}$ calculate the energy consumption because of instructions and data cache, respectively. $I_{miss} \times I_{access\_energy} \times I_{line\_size}$ and $D_{miss} \times D_{access\_energy} \times D_{line\_size}$ is the energy cost of filling data into instruction and data caches from main memory, when a miss occurs. Last two terms calculate the energy cost of the DRAM to respond to cache misses.

In our approach, we remove the first term of the Energy equation $execTime \times CPU_{power}$ because of three reasons: (1) the term $CPU_{power}$ is constant and the term $execTime$ is already being minimized in the first objective, (2) it represents the amount of energy consumed by the CPU and we are optimizing just the performance and energy consumed by the memory subsystem, and (3) in the case of a multi-objective optimization, all the objectives must be as orthogonal as possible, i.e., the term $execTime$ is redundant. 

\subsection{Performance model}

Performance model allows us to obtain the execution time for the cache memory. This model is based on the number of hits and misses in the cache memory subsystem and the time needed to solve them. Equation~\eqref{eq:extime} shows how the execution time is computed. Each component is described below, although a widely and detailed explanation can be found  in~\cite{Janapsatya-2006}. 

\begin{scriptsize} 
\begin{eqnarray}
T=& Icache_{access} \times Icache_{access\_time} + Icache_{miss} \times DRAM_{access\_time} +  \nonumber \\
& Icache_{miss} \times Icache_{line\_size} \times \frac{1}{DRAM\_bwidth} + \nonumber \\
  & Dcache_{access} \times Dcache_{access\_time} + Dcache_{miss} \times DRAM_{access\_time} + \nonumber \\
  &  Dcache_{miss} \times Dcache_{line\_size} \times \frac{1}{DRAM\_bwidth} \label{eq:extime}
\end{eqnarray}
\end{scriptsize}

Terms $Icache_{access}$ and $Dcache_{access}$ correspond to the number of accesses to the instructions and data cache memory, respectively. $Icache_{miss}$ and $Dcache_{miss}$ are the number of cache misses. $Icache_{access\_time}$ and $Dcache_{access\_time}$ represent the time needed to solve each access to the instructions and data cache, respectively. $DRAM_{access\_time}$ is the main memory latency. $Icache_{line\_size}$ and $Dcache_{line\_size}$ are the line or block size for the instructions and data cache, and $DRAM\_{bwidth}$ represents the bandwidth of the DRAM.

Thus, $Icache_{access} \times Icache_{access\_ti\-me}$ and $Dcache\-_{access} \times Dca\-che_{access\_ti\-me}$ compute the total time needed to solve all accesses to the instructions and data cache, respectively. $Ica\-che_{miss} \times DRAM_{access\_ti\-me}$ and $Dca\-che_{miss} \times DRAM_{access\_ti\-me}$ are the total time solving accesses to the main memory, as a consequence of misses in the instructions and data cache. $Ica\-che_{miss} \times Ica\-che_{li\-ne\_si\-ze} \times \frac{1}{DRAM\_bwidth}$ and $Dca\-che_{miss} \times Dca\-che_{li\-ne\_si\-ze} \times \frac{1}{DRAM\_bwidth}$ 
computes the total time needed to fill an instructions or date cache line, when a miss happens. All equations use seconds for time, Watts for power, Joules for energy, bytes for cache line size and bytes/sec for bandwidth.

\section{Register file optimization}
\label{sec:registers}
The first layer of our methodology is the register file optimization. We present a methodology that takes into account the temperature increase due to the accesses that 
happen while a multimedia application is running. Then it evaluates the thermal impact of different spatial distributions of the logical registers. It applies a 
Multi-Objective Evolutionary Algorithm (MOEA) to obtain the optimized solutions, finally proposing the spatial distributions which better reduce the thermal impact. This re-assignment of registers virtually increases the distance \footnote{Distancing means assigning logical registers highly accessed during a program execution to registers that are physically separated in the 
register file.} between registers with higher number of accesses, and it results in a decrease of temperature. 
 
In order to assess our proposal, we have selected a basic register file configuration inspired in two VLIW and ARM architectures. We focus on the $32$ general purpose registers for VLIW and, in the case of ARM, on the $16$ available for all processor modes by replication. Both architectures have different behavior patterns allowing us to analyze and optimize the thermal impact under completely different scenarios. 
  
We have adopted a multigrid design to simulate the physical area of the register file, where the parameters needed by the simulator are the internal structure size, the number of cells and the cell size according to the target architecture. Following the design described in~\cite{Kim-2010}, the register file area is divided into single cells, that will be modified in concordance with the target size. Table~\ref{Table:register_file_size} details the physical measures of the register file on both architectures. Measures are expressed in microns and cells. Columns named ``width'' and ``height'' represent the size of the register file in terms of cells, where a register is $3$ cells wide by $3$ cells high. The next three columns are the size in microns for an individual register and the width and height of the register file.

 \begin{table}[ht]
\renewcommand{\arraystretch}{1.3}
\footnotesize
\centering
\caption{Physical parameters of register file with a register size of $3$ cells high and $90$ cells width.}
\label{Table:register_file_size}
\begin{tabular}{lcccccc} \hline
 & &\multicolumn{2}{c}{\textbf{Cells}} & ($\textbf{width} \times \textbf{height}$)&\multicolumn{2}{c}{\textbf{Register file} ($\mu m$)}\\ \hline
  Arch.& Num. of Registers &Width&Height&Size ($\mu m^2$)&Width&Height \\ \hline
  VLIW&32&90&96&$3\mu m \times 3\mu m$ &270&270 \\
  ARM&16&90&48&$3\mu m \times 3\mu m$ &288&144 \\ \hline
\end{tabular}
\end{table}

\begin{figure}[ht!]
  \centering
  \includegraphics[width=0.75\textwidth]{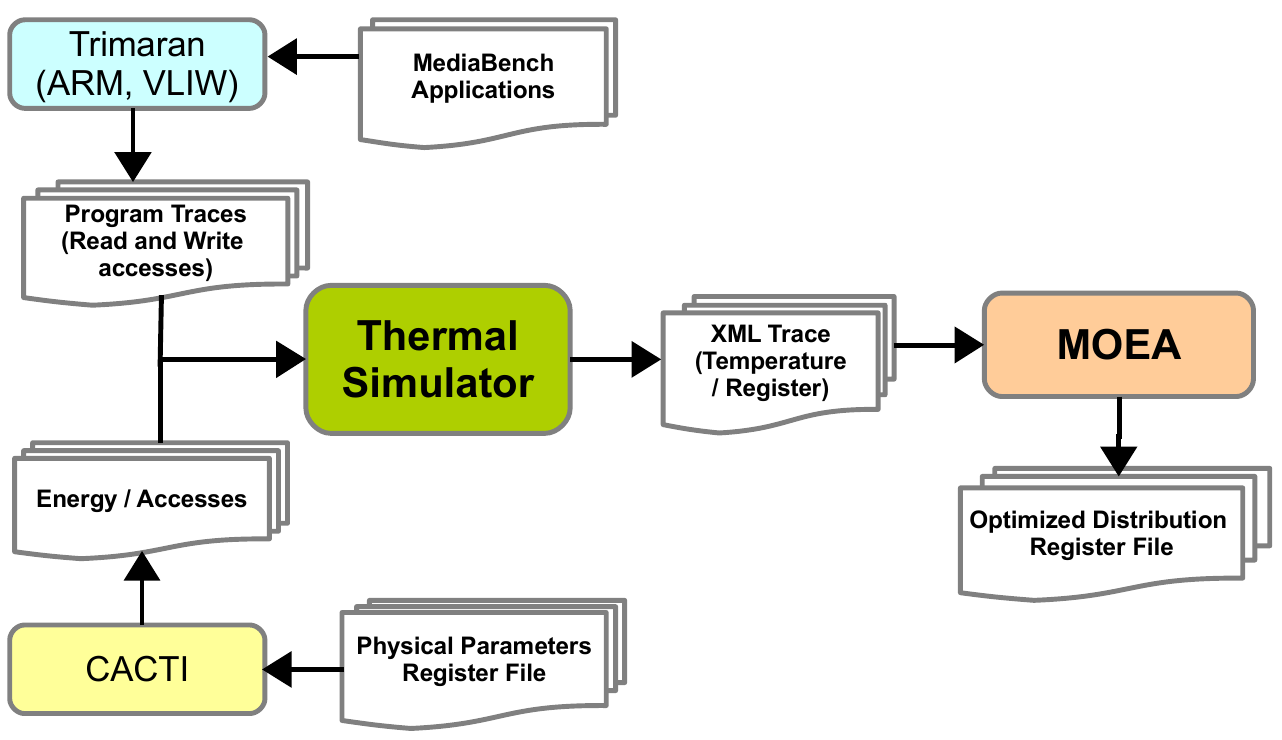}
  \caption{Optimization methodology. (1) Applications are simulated by Trimaran to obtain program traces. (2) CACTI tool is used to compute energy consumption per access. 
  (3) Thermal simulator, designed in Matlab, processes program traces and applies the thermal model defined. (4) Optimization process with NSGA-II as MOEA distributes accesses
  through the register file and minimizes the thermal impact.}
  \label{figure:fig_regfile_metodo} 
\end{figure}

\figurename{~\ref{figure:fig_regfile_metodo}} shows the methodology that we have applied. Firstly, the target application is simulated by Trimaran~\cite{Middha-2002}, a tool widely applied to obtain multiple metrics during running applications. Similarly, energy consumption per access is computed by
CACTI ~\cite{web_cacti}, which is a well-known cache simulator used for estimating energy consumption for different processor structures. These off-line processes must be executed only once. Next, the program trace is processed by the thermal simulator, which generates a customized XML file. This XML file contains one row for each register, which has several associated elements: label, position \textit{x},\textit{y} (inside the design area), width, height and power density, according to the thermal and energy model proposed ($dp=P_\mathrm{reg} /A_\mathrm{reg}$). The XML file is provided as input to the optimization algorithm (MOEA in the Figure), which simulates the internal structure in concordance with the given architecture. The MOEA produces solutions where the geometric register configuration is determined, as well as the register position on each configuration. In this case, the compiler will perform the register re-assignment.   

Our optimization process has just two objective functions: (I) minimize the thermal impact because of the register accesses and the influence of neighbors cells, described by
Equation~\eqref{eq:fitness_reg}, where $c$ is a given configuration, $dp_i$ and $dp_j$ correspond to the power density of the registers $i$ and $j$ and $d_{ij}$ is the 
Euclidean distance between them, and (II) fit the physical viability of the design area that controls $X$ and $Y$ positions inside it. Final human decision is needed to 
select the best solutions, which are analyzed in the next section. Therefore, we have selected the NSGA-II multi-objective algorithm as the required MOEA, following the classical implementation described in \cite{Deb-2002}. Our code is publicly available in \cite{ABSysJECO}. Parameters for the NSGA-II algorithm are specified in Table~\ref{Table:parreg}, and they were adjusted after some preliminary experiments.

\begin{equation}
 \label{eq:fitness_reg}
 f(c) = \sum_\mathrm{i=1}^\mathrm{N_\mathrm{reg}} {\frac{(dp_i \times dp_j)}{d_{ij}}}
\end{equation}

\begin{table}[!ht]
\renewcommand{\arraystretch}{1.3}
\footnotesize
\centering
\caption{Parameters for the NSGA-II algorithm.}
\label{Table:parreg}
\begin{tabular}{lc} \hline
Parameter& Value \\ \hline
Generations & 250 \\
Population Size & 100 \\
Chromosome Length & \textit{Num. registers} \\
Crossover & 0.9 (fixed point) \\
Mutation & 1/\textit{Num. registers} \\\hline
\end{tabular}
\end{table}

We study three possible topologies, $C1$, $C2$ and $C3$ for both VLIW and ARM architectures as shown in Table~\ref{Table:die_configuration}. These topologies have been chosen based on preliminary tests. According to the number of registers, we looked for alternatives to place more or fewer cells into contact with the external part of the register file, which allows us to better study the thermal behavior. Regarding the target applications, we have selected a subset of the multimedia benchmark Mediabench~\cite{Mediabench} (\textit{epic}, \textit{unepic}, 
\textit{cjpeg}, \textit{djpeg}, \textit{mpegdec}, \textit{mpegenc}, \textit{gsmdecode}, \textit{gsmenconde}, \textit{rawcaudio}, \textit{rawdaudio}), widely used in the scientific 
community to increase the register file traffic.

\begin{figure}[ht]
  \centering
  \includegraphics[width=0.95\textwidth]{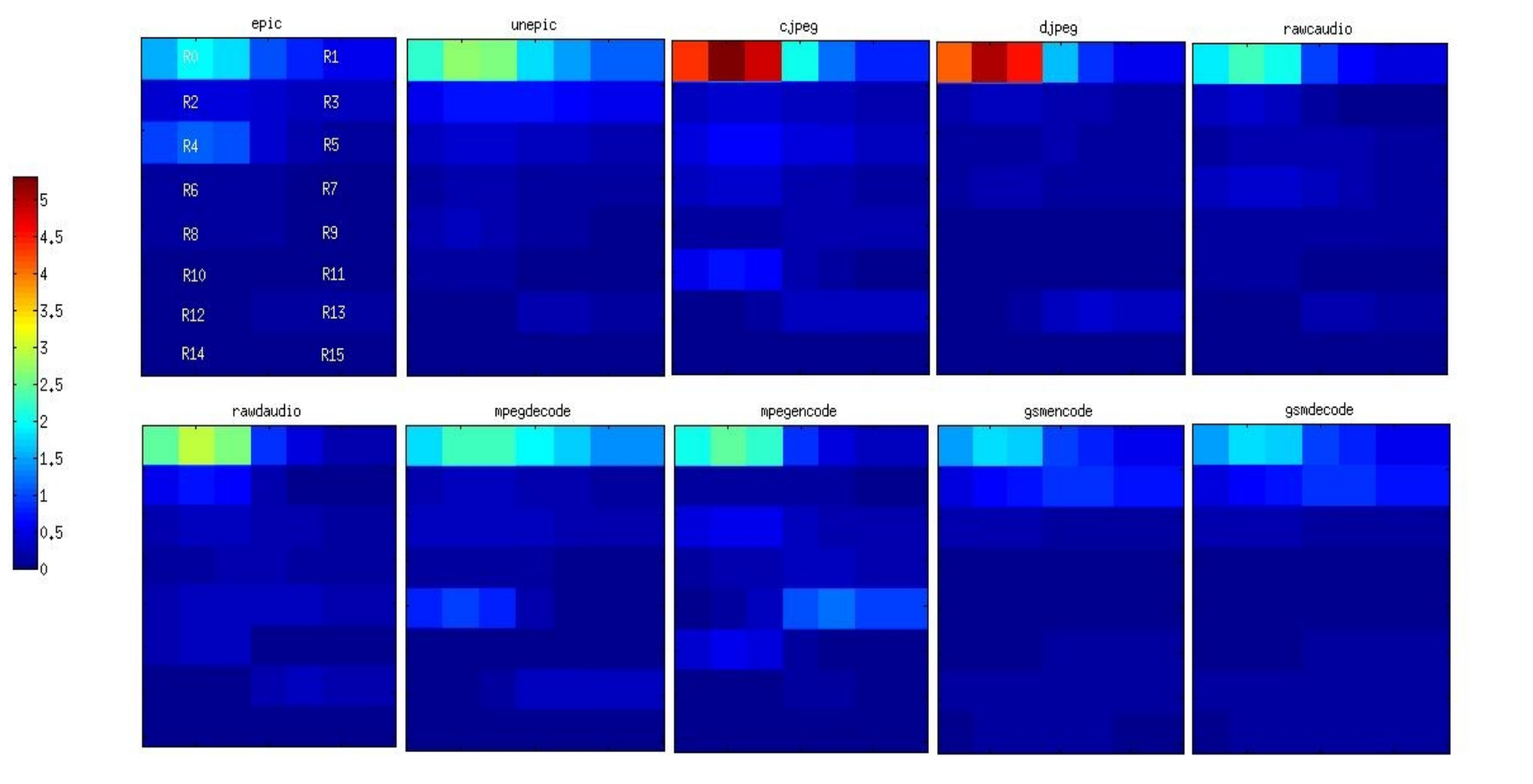}
  \caption{Thermal impact because of accesses to general purpose registers in a typical ARM architecture before the optimization. $C3$ topology has $2$ rows and $8$ columns, although in better shape, graphs are shown in $8 \times 2$ format.}
  \label{figure:arm_c3} 
\end{figure}

Experimental results show that the thermal impact in the VLIW architecture is not significant. The maximum increase in temperature for VLIW is $0.4319 ^\circ C$ in $C1$ and $C2$ configurations and $0.4318 ^\circ C$ in $C3$. In the case of ARM this increase is $5.3044 ^\circ C$ in $C1$ and $C2$, and $5.3036 ^\circ C$ in $C3$. So, we focus on the ARM architecture, where the temperature increase presents some interesting values that we consider to be optimized.~\figurename{~\ref{figure:arm_c3}} shows the thermal impact on configuration $C3$ before the optimization with a $2$ rows $\times$ $8$ columns topology, as an example. For the sake of the space, Figures display $8$ rows and $2$ columns.

\begin{table}[!ht]
\renewcommand{\arraystretch}{1.3}
\footnotesize
\centering
\caption{Physical description of register file.}
\label{Table:die_configuration}
\begin{tabular}{cccccccc} \hline
\multicolumn{4}{c}{\textbf{VLIW Architecture} } & \multicolumn{4}{c}{\textbf{ARM Architecture}}\\ \hline
Conf. & $Rows \times Columns$ & Width & Hight & Conf. & $Rows \times Columns$ & Width & Hight\\ \hline
 C1 & $32 \times 1$ & 90 & 96 & C1 & $16 \times 1$ & 90 & 48 \\  
 C2 & $16  \times 2$ & 180 & 48 & C2 & $8 \times 2$  & 180 & 24 \\  
 C3 & $4 \times 8 $& 720 & 12 & C3 & $2 \times 8$  & 360 & 6\\ \hline
\end{tabular}
\end{table}

Table~\ref{Table:resumen_arm} shows the improvement percentage for all applications after the optimization. ~\figurename{~\ref{figure_arm_g3} shows the case of $C3$ configuration graphically, after the optimization process, which presents the best behavior among the studied topologies.

\begin{table}[!ht]
\renewcommand{\arraystretch}{1.3}
\centering
\caption{ARM. Improvement percentage with respect to the average and maximum temperature.}
\label{Table:resumen_arm}
\begin{tabular}{ccccccccccc}\hline
&\multicolumn{2}{c}{\textbf{epic}}&\multicolumn{2}{c}{\textbf{unepic}}& \multicolumn{2}{c}{\textbf{cjpeg}}& \multicolumn{2}{c}{\textbf{djpeg}}&
\multicolumn{2}{c}{\textbf{gsmdec}} \\ \hline 
 &Avg&Max&Avg&Max&Avg&Max&Avg&Max&Avg&Max \\ \hline
C1&3.87&2.14&5.55&3.10&1.84&1.01&1.22&0.67&4.12&2.28 \\ 
C2&3.74&2.15&5.51&3.10&1.81&1.01&1.21&0.67&4.03&2.29 \\ 
C3&3.78&2.15&5.54&3.11&1.83&1.01&1.22&0.67&4.13&2.29 \\ \hline
&\multicolumn{2}{c}{\textbf{gsmenc}}& \multicolumn{2}{c}{\textbf{rawc}}& \multicolumn{2}{c}{\textbf{rawd}}&\multicolumn{2}{c}{\textbf{mpegdec}}&
\multicolumn{2}{c}{\textbf{mpegenc}}\\ \hline 
&Avg&Max&Avg&Max&Avg&Max&Avg&Max&Avg&Max \\ \hline 
C1&4.15&2.31&2.61&1.44&1.00&0.55&7.75&5.36&1.56&0.85 \\ 
C2&4.07&2.31&2.57&1.44&1.00&0.55&7.68&5.36&1.53&0.86 \\ 
C3&4.17&2.32&2.59&1.44&1.01&0.55&7.73&5.36&1.55&0.86 \\ \hline
\end{tabular}
\end{table}

\begin{figure}[ht]
  \centering
  \includegraphics[width=0.95\textwidth]{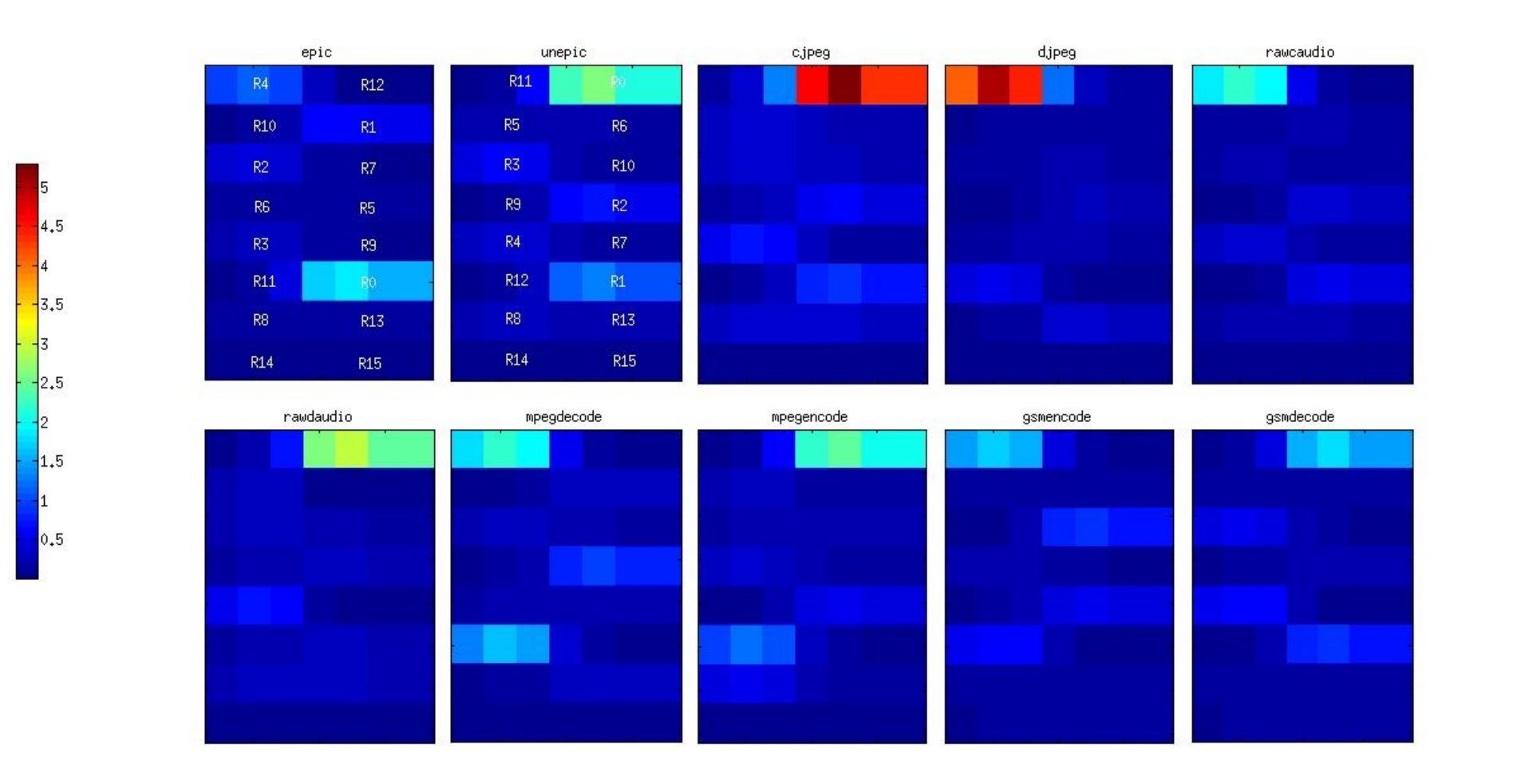}
  \caption{Thermal map for the $C3$ topology in ARM architecture, after the multi-objective optimization process. Registers with lower thermal impact are placed next to others with higher thermal impact, so the temperature in the whole structure decreases for all applications, as Table~\ref{Table:resumen_arm} shows.
  }
  \label{figure_arm_g3} 
\end{figure}

As seen in both Figures, \textit{hot spots} have been moved in the majority of the cases. However, in some of the benchmarks like \textit{cjpeg} and \textit{djpeg}, the influence of the most accessed register in their neighborhood is so high that the adjacent registers are highly affected by its temperature increase. However, the average temperature values shows that, despite the decrease in terms of temperature is not relevant, the proposed method is able to reduce the thermal impact of the register file on the processor temperature in all the architectures, configurations and multimedia applications addressed. As shown in~\figurename{~\ref{figure_arm_g3}}, registers highly accessed are placed in the outside borders of the register file, in order to facilitate power dissipation.  

\begin{table}[ht]
\renewcommand{\arraystretch}{1.3}
\centering
\caption{t-Student test for each application and configuration with respect to the average of the maximum non-optimized temperature in the ARM architecture.}
\label{Table:statreg}
\begin{tabular}{lrrrrrr}\hline
Application&\multicolumn{2}{c}{\textbf{C1}}&\multicolumn{2}{c}{\textbf{C2}}& \multicolumn{2}{c}{\textbf{C3}} \\ \hline 
 &T&\textit{p-value}&T&\textit{p-value}&T&\textit{p-value} \\ \hline
epic&0.2161&0.5841&-0.3366&0.3705&0.3335&0.6283\\ 
unepic&0.231&0.5898&-0.0638&0.475&0.2802&0.6084\\
cjpeg&0.001&0.5004&-0.0554&0.4783&0.031&0.5122\\
djpeg&0.0148&0.5058&-0.0337&0.4868&0.0261&0.5102\\
gsmdec&0.2933&0.6133&-0.1101&0.4569&0.3326&0.628\\
gsmenc&0.2969&0.6147&-0.1122&0.4561&0.3358&0.6292\\
rawcaudio&0.0354&0.5139&-0.1318&0.4484&0.0526&0.5206\\
rawdaudio&0.0325&0.5127&-0.0098&0.4262&0.0158&0.5062\\
mpegdec&0.1721&0.5672&-0.1808&0.4295&0.2144&0.5835\\
mpegenc&-0.1289&0.4496&-0.0411&0.4839&0.0009&0.5003\\
\hline
\end{tabular}
\end{table}

With the aim to verify the statistical relevance of the data obtained, we have performed the statistical \textit{t-Student} test. We have used the statistical software \textit{R}~\cite{r2014} to perform this test and compute the \textit{p-value}. The \textit{p-value} is a statistical measure that allows determining if our results are significant. \textit{T} represents the statistical value used to make the test. Regarding the freedom degree, which is the number of freely chosen values in a sample which allow reaching a value, we consider the number of registers. Hence, for the problem at hand we verify statistically that the decrease of temperature obtained after the optimization process is not relevant in both the ARM and VLIW architectures.

Table~\ref{Table:statreg} shows the statistical results for the ARM architecture in the proposed topologies and all multimedia applications. For each topology and optimized solution, the temperature increase of its registers is compared to the average of the maximum temperature in the non-optimized configuration. The \textit{p-value} is higher than $0.05$ in all cases, thus we conclude that the decrease of temperature obtained is not relevant. The same conclusion is applicable to the VLIW architecture, where we have obtained \textit{p-value} values over $0.05$ in all cases, too. We have omitted the VLIW table for the sake of space.

\section{Cache memory optimization}
\label{sec:cache}
The second layer of our methodology is the cache memory, as previously shown in~\figurename{~\ref{figure:method_layer}}. We propose an optimization approach which is able to determine cache configurations for multimedia embedded systems and require less execution time and energy consumption.

\begin{figure}[ht]
  \centering
  \includegraphics[width=0.85\textwidth]{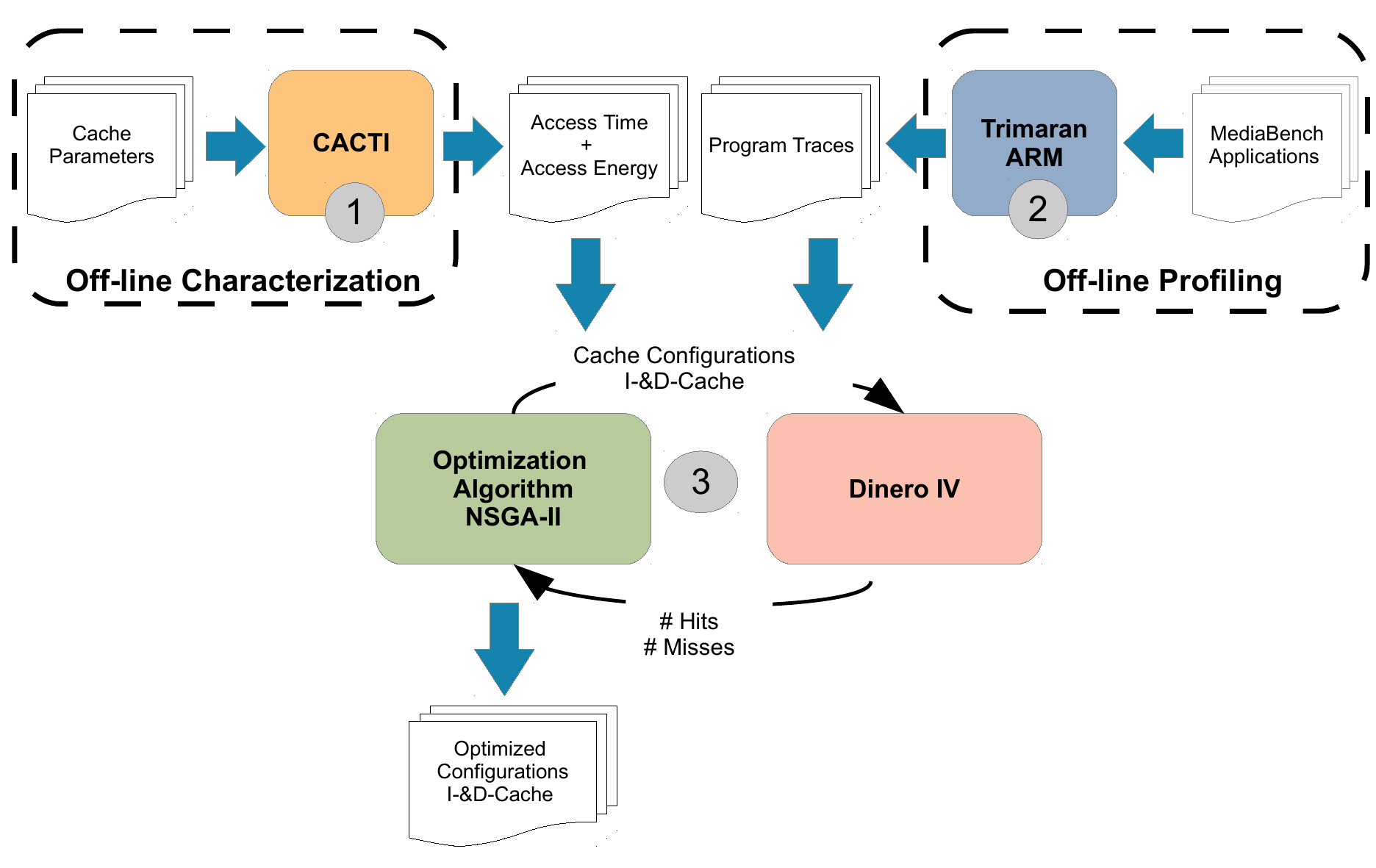}
  \caption{Optimization process: (1) cache characterization, (2) application profiling and (3) cache optimization.} 
  \label{figure:method}
\end{figure}

As seen in ~\figurename{~\ref{figure:method}}, this layer is divided into two off-line phases (labeled as 1 and 2) and a third phase devoted to optimization (labeled as 3). 
Firstly, the off-line phases are executed just once before the optimization. Next, the optimization process uses as input the results of the previous two phases. The cache 
characterization phase is performed by CACTI, which computes access times and energy consumed by the addressed structures. These values are necessary to calculate the objective functions while the evolutionary algorithm is running. The application profiling phase is carried out using Trimaran, which compiles all cache memory accesses into program traces. In the third phase, the 
NSGA-II optimization algorithm  evaluates each candidate solution with the help of Dinero IV \cite{DineroIV}. In this case, we have modified our implementation of NSGA-II in order to be able to evaluate solutions using an external program like Dinero IV. Dinero IV is a cache simulator, which given a program trace as 
input, computes the number of hits and misses of memory accesses. These metrics multiplied by either the access delays and the energy per access previously given by 
CACTI, provide the optimization algorithm a quality measure for each individual under evaluation. Table~\ref{Table:parcacheNSGAII} shows the parameter values used to configure the NSGA-II algorithm. These values were selected after some preliminary experiments.

\begin{table}[!ht]
\renewcommand{\arraystretch}{1.3}
\footnotesize
\centering
\caption{Parameters for the NSGA-II algorithm.}
\label{Table:parcacheNSGAII}
\begin{tabular}{lc} \hline
Parameter& Value \\ \hline
Generations & 250 \\
Population Size & 100 \\
Chromosome Length & 11 \\
Crossover & 0.9 (fixed point) \\
Mutation & 1/11 \\\hline
\end{tabular}
\end{table}

Given the different values that each one of the cache configuration parameters may take, a very high number of combinations can be generated. \figurename{~\ref{figure:arboles}} shows the design space defined by the set of cache parameters (cache size, block size, associativity, 
replacement algorithm and prefetch algorithm for both the instruction and data caches, and also write policy for data cache) and their possible values. 

\begin{figure}[ht]
	\centering
 	\includegraphics[width=0.85\textwidth]{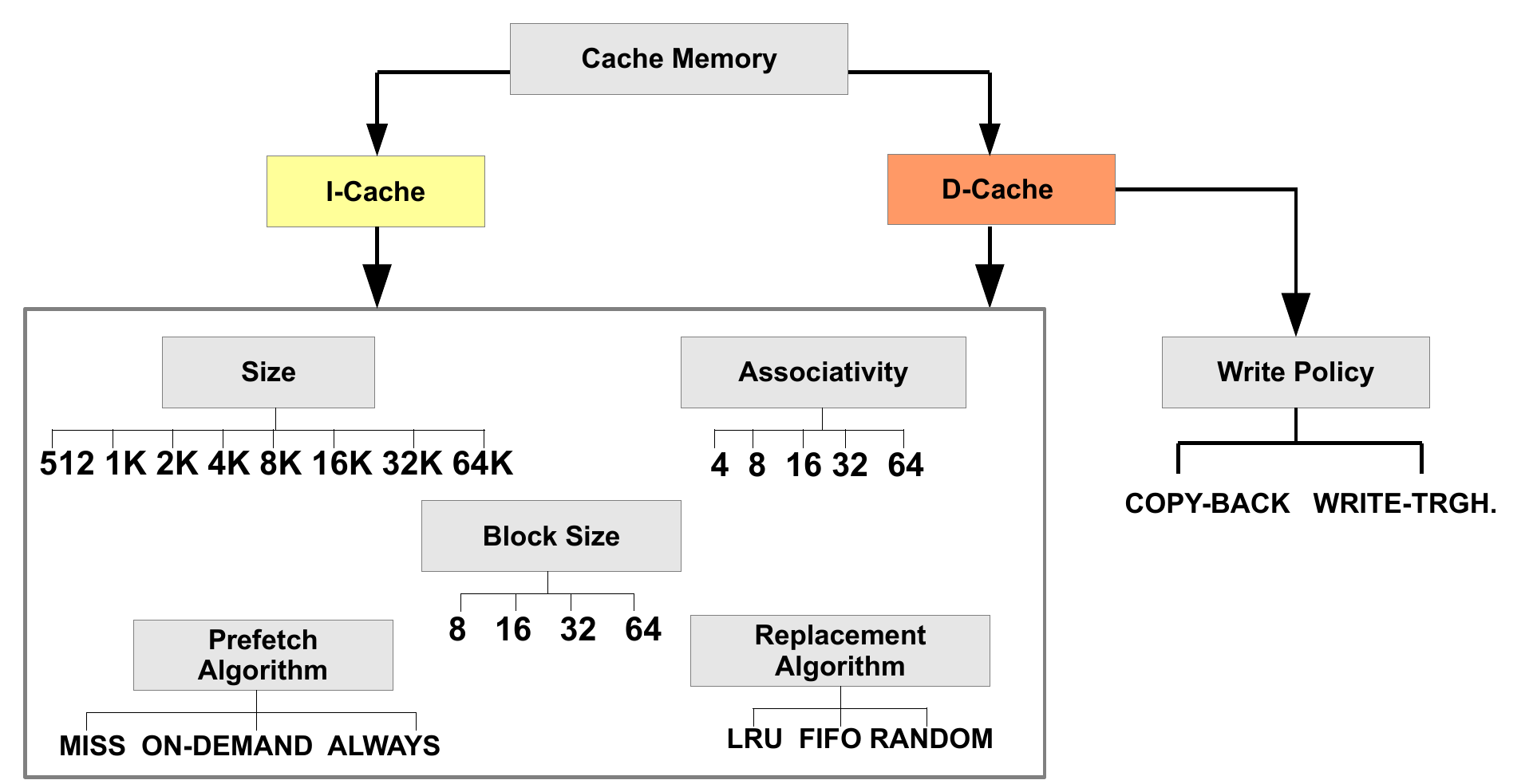}
 	\caption{Taxonomy for a cache configuration design space. Both instructions and data caches, labeled as I-Cache and D-Cache, must be customized with their corresponding values.}
  \label{figure:arboles} 
\end{figure}

The NSGA-II optimization algorithm deals again with the objectives corresponding to execution time and energy consumption, applying the previously defined equations 
~\eqref{eq:extime} and~\eqref{eq:energy1}.

As we will describe in the next section, the resulting cache configurations are compared with three baseline cache configurations. However, our 
methodology might use any other cache configuration as a reference baseline.


Experimental results are based on the ARM architecture, particularly ARM\-920T~\cite{ARM9}. Again, we have selected a set of applications from Mediabench~\cite{Mediabench}. In this case, we have also considered \textit{pegwitdec} and \textit{pegwitenc} in addition to those benchmarks  previously mentioned, all of them with their standard inputs. It is well-known that multimedia applications from Mediabench increase the pressure on the cache memory due to the intrinsic nature of data they manage.

Every application has been run for $7.5 \times 10^7$ instructions to reach a balance between the total execution time, the size of the generated program traces and a proper number of instructions.

The optimization process driven by NSGA-II is able to find cache configurations that presents an average improvement of execution time and energy consumption of $33.9\%$ and $71.8\%$ respectively, taking a baseline as a reference. 
Table~\ref{Table:baselines} shows the three baseline configurations used in order to test our methodology. The first baseline is similar to the L1 cache of the first core in the GP2X video games console. The second and third one are implemented in Cortex-A9 and Cortex-A15 processor.  

\begin{table*}[ht!]
\centering
\tiny
\renewcommand{\arraystretch}{1.3}
\caption{Baseline cache configurations. Cache size (Size), block size (BS), associativity (A), replacement algorithm (RA), prefetch algorithm (PA) and write policy (WP) are shown for the three configurations.}
\label{Table:baselines}
\begin{tabular}{ccccccccccc} \hline
\multicolumn{5}{c}{\textbf{Instructions Cache}} &\multicolumn{6}{c}{\textbf{Data Cache}}\\ \hline \hline
\textbf{Size}&\textbf{BS}&\textbf{A}&\textbf{RA}&\textbf{PA} &\textbf{Size}&\textbf{LS}&\textbf{A}&\textbf{RA}&\textbf{PA} &\textbf{WP} \\ \hline
16 KB & 32 B & 4 & LRU & On-Dem. & 16 KB & 32 B & 4 & LRU & On-Dem. & Copy-Back \\  \hline
32 KB & 64 B & 4 & Random & Always & 32 KB & 64 B & 4 & Random & Always & Copy-Back \\  \hline
32 KB & 64 B & 2 & LRU & Always & 32 KB & 64 B & 2 & LRU & Always & Copy-Back \\  \hline
\end{tabular}
\end{table*}

The results we obtained are shown in Table~\ref{table:appleserie}, where the comparison with the three baselines is made. It is important to highlight that the Pareto set\footnote{Set of solutions returned by a MOEA, which must take an uniform distribution on the objective space and be as close 
as possible to the Pareto Optimal Front (POF). The POF represents the set of the best possible solutions for a given problem.} of some applications regarding the Baseline 1 includes solutions with negative improvement percentage for one of the objectives. These values not only reduce the average improvement in an application but also the total average improvement. We have not removed or penalized these solutions because, on one hand, they are not unfeasible solutions but they are solutions with performance worse than the baseline reference and, on the other hand, they can illustrate the decision maker how far a configuration can perform if one of the objectives is relaxed. As a consequence, the average percentages corresponding to Baseline 1 would be increased to $34.98\%$ and $79.34\%$ in terms of execution time and energy consumption, respectively. These percentages would be higher, if we define a quality standard for the solutions, as part of the human decision-making phase.

In relation to the other two baselines, corresponding to the \textit{SoC Apple AX} series included in different Apple devices like Cortex-A9 (iPad 2, iPod-touch o iApple-TV) and Cortex-A15 (iPhone 5, iPhone 5S), the results are even better than in the other baseline.  The optimization process finds cache configurations that improve on average $49.37 \%$, $93.24\%$ and $44.84\%$, $93.82\%$ with respect to baselines 2 and 3 in terms of execution time and energy consumption, respectively, as shown in Table \ref{table:appleserie}.

\begin{table*}[!ht]
\centering
\tiny
\caption{Average improvement percentage of solutions belonging to Pareto set vs. the three  baselines under study.}
\begin{tabular}{l|cc|cc|cc}
\hline
\multirow{2}{*}{}&\multicolumn{2}{c}{\textbf{Baseline 1}} &\multicolumn{2}{c}{\textbf{Baseline 2}} &\multicolumn{2}{c}{\textbf{Baseline 3}}\\
 \textbf{App.}  &\textbf{Ex. T. (\%)} & \textbf{Energy  (\%)} &\textbf{Ex. T. (\%)} & \textbf{Energy  (\%)} &\textbf{Ex. T.  (\%)} & \textbf{Energy  (\%)} \\ \hline
epic &44.9&76.3 & 55.77 & 91.94 & 51.35 &92.63\\
unepic &33.4&26.7&42.58&93.82&37.05&94.39 \\
gsmdec &31.4&84.3&36.25&94.24&30.39&94.74 \\
gsmenc &19.0&83.6&33.76&94.07&31.15&94.36 \\
cjpeg &27.2&71.1&52.63&91.04&47.82&91.78 \\
djpeg &16.5&72.4&44.86&91.06&40.02&92.01 \\
pegwitdec &27.1&83.9&51.31&95.15&46.77&95.44 \\
pegwitenc &35.8&84.8 &53.76&95.96&49.51&96.20 \\
rawcaudio &48.8&84.8&59.45&94.39&55.18&94.83 \\
rawdaudio &48.1&78.4&60.02&91.94&55.81&92.58 \\
mpegdec &37.9&71.2&52.40&90.68&48.48&91.77 \\
mpegenc &37.9&43.9&49.65&94.60&44.55&95.06 \\ \hline
\textbf{Average} &\textbf{33.9}&\textbf{71.8}&\textbf{49.37}&\textbf{93.24}&\textbf{44.84}&\textbf{93.82} \\ \hline
\end{tabular}
\label{table:appleserie}
\end{table*}

Figures~\ref{nsgaii_1} and~\ref{nsgaii_2} display the cache configurations in the form of Pareto fronts where the axis correspond to the execution time and energy consumption objective functions. As seen in the plots, in all applications the solutions are uniformly distributed in the objective space. Some applications such as \textit{epic}, \textit{unepic}, \textit{gsmdec} and \textit{gsmenc} present a greater uniformity than the rest, but all of them provide a high number of solutions. 

Given that the cache configurations provided by the optimization algorithm present different performances, they all should be provided to the cache designer. The expert will decide the best solution to fit the requirements of the target system. In this context, our method simplifies this selection process and provides a set of good solutions to the system designers.


\begin{figure*}[!t]
\centering
  \includegraphics [width=5in] {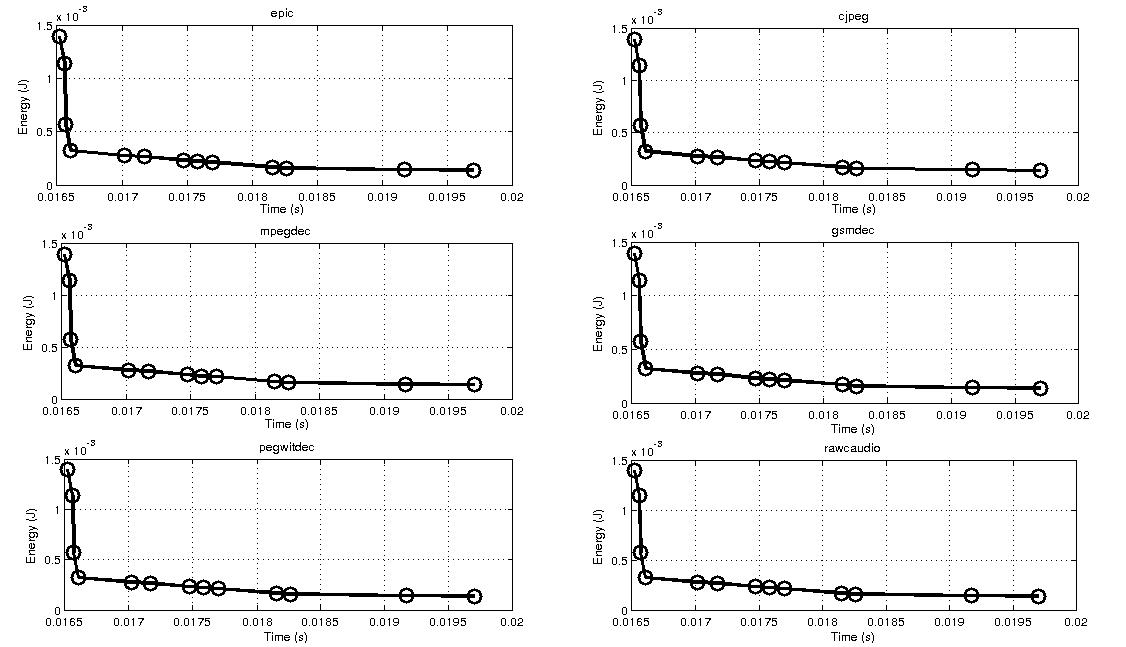}
  \caption{Pareto front representation for NSGA-II for epic, cjpeg, mpegdec, gsmdec, pegwitdec and rawcaudio applications. Results show that solutions are uniformly distributed in the design space and cover a wide 
  region.}
  \label{nsgaii_1}
\end{figure*}
\begin{figure*}[!t]
\centering
  \includegraphics [width=5.1in] {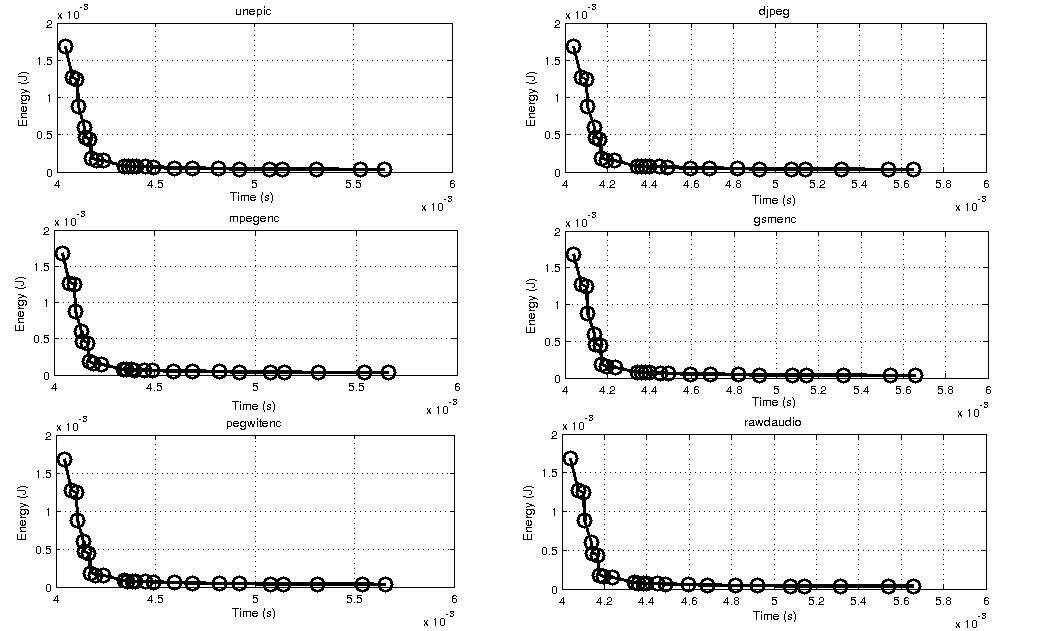}
  \caption{Pareto front representation for NSGA-II for unepic, djpeg, mpegenc, gsmenc, pegwitenc and rawdaudio applications. Results show that solutions are uniformly distributed in the design space and cover a wide 
  region.}
  \label{nsgaii_2}
\end{figure*}

With the aim to measure the relevance of results presented, we have computed the statistical \textit{t-Student} test as in the previous section. The obtained results are shown in Table~\ref{Table:p_value1}, 
where \textit{FD} is the freedom degree.

\begin{table}[!t]
\centering
\scriptsize
\renewcommand{\arraystretch}{1.8}
\setlength\belowcaptionskip{5pt}
\caption{t-Student test for the Pareto set of the applications with respect to a baseline cache.}
\label{Table:p_value1}
\begin{tabular}{llrrrr} \hline
&\multicolumn{3}{c}{\textbf{Time}}&\multicolumn{2}{c}{\textbf{Energy}}\\ \hline
\multicolumn{1}{c}{\textbf{Aplication}} &  \multicolumn{1}{c}{\textbf{FD}} &  \multicolumn{1}{c}{\textbf{T}}&
 \multicolumn{1}{c}{\textbf{\textit{p-value}}} &  \multicolumn{1}{c}{\textbf{T}}&  \multicolumn{1}{c}{\textbf{\textit{p-value}}}\\ \hline
cjpeg&24&-5.47&6.42E-006&-15.87&1.59E-014\\ \hline 
djpeg&19&-3.58&1.01E-003&-14.48&5.10E-009\\ \hline 
epic&12&-50.71&1.13E-012&-11.61&3.48E-005\\ \hline 
unepic&26&-23.79&2.20E-016&-1.23&1.15E-001\\ \hline 
unepic&23&-22.09&2.20E-016&-5.52&6.41E-003\\ \hline 
gsmdec&16&-17.66&3.23E-012&-32.69&2.20E-016\\ \hline 
gsmenc&14&-7.59&1.27E-003&-28.48&4.28E-011\\ \hline 
mpegdec&16&-25.96&8.30E-012&-11.38&2.20E-006\\ \hline 
mpegenc&15&-24.21&9.77E-011&-1.60&6.50E-002\\ \hline 
mpegenc&13&-22.93&3.36E-009&-17.61&9.38E-008\\ \hline 
pegwitdec&11&-4.53&4.31E-004&-22.55&7.35E-008\\ \hline 
pegwitenc&12&-13.09&9.15E-006&-25.81&3.49E-009\\ \hline 
rawcaudio&12&-63.21&2.20E-016&-21.43&3.11E-008\\ \hline 
rawdaudio&12&-54.49&4.81E-013&-13.65&5.70E-006\\ \hline 
\end{tabular}
\end{table}

In this case, the freedom degree corresponds to the number of solutions of the Pareto set in all applications.
We observe in the table that the \textit{p-value} is far lower than $0.05$ for all  applications with respect to the performance and the energy consumption. As a result of the statistical test, we can say the results obtained with the proposed optimization methodology are relevant.

\section{Dynamic memory management optimization}
\label{sec:main_memory}
The third layer of our methodology consists on an optimization framework based on GE and static profiling of applications to improve the dynamic memory manager (DMM) for 
multimedia applications, which have high dependence of dynamic memory. This is a non-intrusive method that allows to automatically evaluate complex implementations of DMMs. 

In order to evaluate our proposal, we have selected six memory intensive applications: \textit{hmmer}, \textit{dealII}, \textit{soplex}, \textit{calculix}, \textit{gcc} and \textit{perl}. In addition, we have taken the Lea DMM (implemented in GNU/Linux systems) and the Kingsley DMM (implemented in Windows systems) as references to normalize the performance of the results. In fact, we analyzed the execution time, memory footprint and temperature of the memory, according to the thermal model proposed in~\cite{Brooks-2007}, and the energy consumption following the energy model developed in~\cite{Qa2009} in a set of preliminary experiments. As a result, we found that the Lea DMM has a high impact on the performance and in the memory footprint, circa $43.25\%$ and $22.9\%$, respectively. On the other hand its influence is not significant with respect to the temperature and energy consumption, which is $0.0006\%$ and  $0.48\%$ on average. The Kingsley DMM presented a similar behavior, so we decided to use the first two metrics of performance and footprint.

\begin{figure}[!t]
\centering
\includegraphics[width=0.85\textwidth]{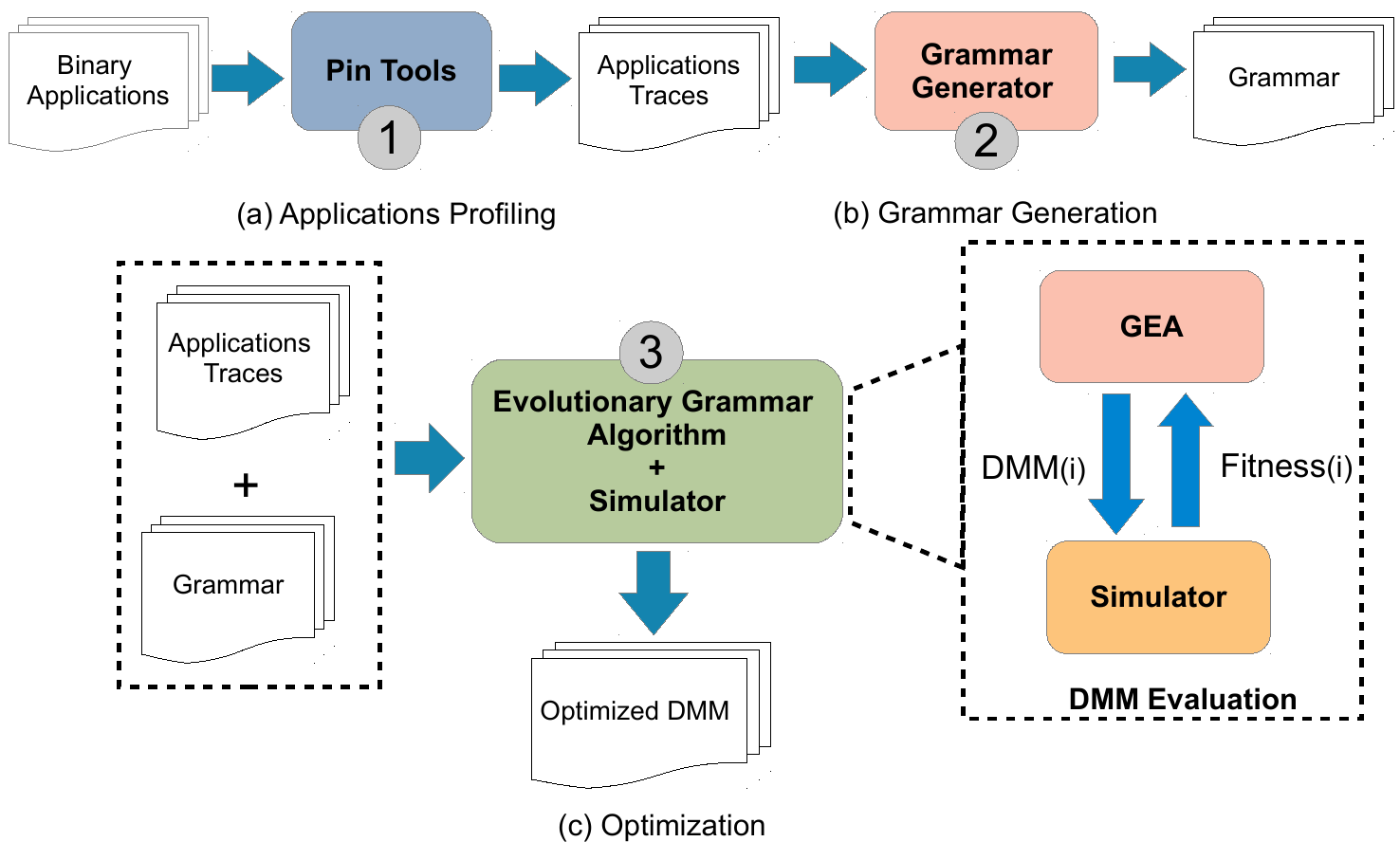}
\caption{DMM optimization: (1) obtains applications traces, through Pin as the instrumentation tool; (2) designs the 
customized grammar according to the application trace with and (3) runs optimization algorithm with GE, which generates a customized DMM and the DMM simulator is 
called to obtain the necessary metrics to evaluate it.}
\label{figure:met_opt_dmm}
\end{figure}

Similarly to the other layers, in this case our methodology is divided into three phases, as shown in~\figurename{~\ref{figure:met_opt_dmm}}, a detailed view of 3rd layer 
in~\figurename{~\ref{figure:method_layer}}. The first phase obtains application traces with the Pin instrumentation tool~\cite{Luk2005}. The second phase analyzes the target 
application trace and creates the grammar that best fits with the application patterns. Finally, the optimization algorithm based on GE is run, coupled to a DMM 
simulator~\cite{Risco-2011}, which will collect the metrics needed to obtain the quality for each DMM evaluated. These metrics are the number of memory accesses, memory usage, 
de/allocations, splittings and coalescings. The execution time devoted to the DMM is calculated as the computational complexity given that the system uses simulation time 
instead of real time. 

It is important to explain that the grammar of the GE algorithm will be used to compose the custom DMMs. Therefore, we decided to produce a different grammar for each one of the target applications in order to reduce the search space and, therefore, improve the optimization process. For the sake of space we do not describe here the grammar, but the interested reader may obtain detailed information about these kind of grammars in \cite{RiscoMartin2009b}. The parameters of the GE algorithm are detailed in Table~\ref{Table:par_dmm_gen}, and they were adjusted after some preliminary experiments.

\begin{table}[!ht]
\renewcommand{\arraystretch}{1.3}
\centering
\setlength\belowcaptionskip{5pt}
\caption{Parameters for the GE algorithm.} 
\label{Table:par_dmm_gen} 
\begin{footnotesize}
\begin{tabular}{l l}
\hline
Parameter & Value \\ \hline
Generations & 250 \\
Population Size & 100 \\ 
Chromosome Length & 200 \\
Selection mechanism & Tournament (size=2)\\
Crossover & 0.8 (fixed point)\\
Mutation & 0.02 \\
Maximum\textit{wraps} & 3 \\ \hline
\end{tabular}
\end{footnotesize}
\end{table}

Besides, in our preliminary experiments, we have also verified that the behavior of the applications is similar among different executions. Thus, each target application must be executed just once to obtain the profiling report, which can be used to evaluate different DMMs.

Although GE performs well and does not require high amounts of memory, it tends to fall into a local optimum if it is not correctly set up \cite{Castle2010}. To address this challenge, we have been successfully using premature convergence prevention techniques \cite{Zapater2016}, that are next explained.

Premature convergence of a Genetic Algorithm arises when the chromosomes of some high rated individuals quickly dominate the population, reducing diversity, and constraining it to converge to a local optimum. Premature convergence is one of the major shortcomings when trying to model low variability magnitudes by using GE techniques. To overcome the lack of variety in the population, work by Melikhov et al. \cite{Melikhov1996} proposes the usage of Social Disaster Techniques (SDT). This technique is based on monitoring the population to find local optima, and apply an operator:

\begin{enumerate}
\item \emph{Packing}: all individuals having the same fitness value except one are fully randomized.
\item \emph{Judgment day}: only the fittest individual survives while the remaining are fully randomized.
\end{enumerate}

Work by Rocha et al. \cite{Rocha1999} proposes the usage of Random Offspring Generation (ROG) to prevent the crossover of two individuals with equal genotype, as this would result in the offspring being equal to the parents. Individuals are tested before crossover and, if equal, then one offspring (1-RO) or both of them (2-RO) are randomly generated. 

Both previous solutions have shown important benefits in classical Genetic Algorithms problems. In our work, we use these techniques to improve the convergence time of our solutions, with excellent results. Otherwise, without these enhancements, the standard implementation of GE does not perform well and is not able to find good solutions in a reasonable amount of time, as we have already tested.


The GE optimization algorithm is implemented as an improvement of a Genetic Algorithm with integer chromosomes where the grammar decodification is included in the evolutionary process. As in the previous cases, we have published the code of this algorithm in our JECO library \cite{ABSysJECO}. Regarding the quality of the solutions, the algorithm uses the objective function described in Equation~\ref{eq:coste_dmm_opt} to select the best possible DMM among the candidate solutions. The 
execution time ($T$) and the memory use ($M$) have equal weight and they are normalized to the corresponding Kingsley and Lea DMMs, which are considered the fastest and more
efficient (in terms of memory footprint) DMMs, respectively. Thus, $T$ and $M$ are the execution time and the memory use for the DMM that is currently being evaluated; $T_{Kng}$ 
and $M_{Lea}$ are the execution time and the memory use for Kingsley and Lea DMMs, respectively.

\begin{equation}
\label{eq:coste_dmm_opt}
 F = 0.5 \times \frac{T}{T_{Kng}} + 0.5 \times \frac{M}{M_{Lea}}
\end{equation}

This methodology has been tested with six memory-intensive applications from the SPEC bechmarks~\cite{spec-2013} using standard inputs: \textit{hmmer}, \textit{dealII}, \textit{soplex}, \textit{calculix}, \textit{gcc} and \textit{perl}. We have compared the DMMs obtained by the GE algorithm (GEA) with five different general purpose DMMs: \textit{Kingsley} (KNG), \textit{Doug Lea} (LEA), a \textit{buddy system} based on the Fibonacci algorithm (FIB), a list of \textit{10 segregated free-lists} (S10) and an \textit{exact segregated free list} (EXA). 

Table~\ref{Table:average_result_dmm} shows the average improvement percentage of GEA versus the general purpose DMMs we compare with.  As seen in the table, this method reduces the objective function of weighted execution time and memory use by $59.27\%$ on average. Besides, all the comparisons are positive for the GEA, therefore obtaining better results than the general purpose DMMs.


Hence, our methodology is able to automatically design customized DMMs according to a given application in a non-intrusive way, improving the performance of standard DMMs.

\begin{table*}[!t]
\renewcommand{\arraystretch}{1.3}
\centering
\footnotesize
\setlength\belowcaptionskip{5pt}
\caption{Average improvement percentage. GEA vs. KNG, LEA, FIB, S10 y EXA.} 
\label{Table:average_result_dmm} 
\begin{tabular}{l l l l l l l}
\hline
& KNG & LEA & FIB & S10 & EXA & Average \\ \hline
Obj. value = $100 \times \frac{F_* - F_{GEA}}{F_*}$ & 9.13\% & 62.52\% & 51.81\% & 86.88\% & 86\% & 59.27\% \\
Performance = $100 \times \frac{T_* - T_{GEA}}{T_*}$ & 1.17\% & 72.44\% & 62.62\% & 85.74\% & 90.78\% & 62.55\% \\
Memory = $100 \times \frac{M_* - M_{GEA}}{M_*}$ & 16.03\% & 23.14\% & 15.08\% & 38.88\% & 59.96\% & 30.62\% \\ \hline
\end{tabular}
\end{table*}

We have tested the relevance of the results obtained performing a statistical analysis with the Wilcoxon's matched pair test. For the GEA DMM, the execution time was compared to Kingsley and the memory use was compared to LEA, which are the faster and more efficient respectively using all the applications under study. The results of these tests are shown in Table ~\ref{Table:dmm_statistics}.

The tests performed confirm the conclusions previously mentioned. The test between GEA and Kingsley in performance gives a \textit{p-value} of $0.395$, which indicates that we cannot ensure the samples are different. In fact, the results obtained by GEA in terms of performance are close to Kingsley. Moreover, the Wilcoxon's test with respect to the memory use provides us a $\textit{p-value}$ of $0.142$, which indicates that results obtained are very similar. Again, this is obtained because the use of memory of GEA is not very different from the memory consumption of LEA.


\begin{table*}[!t]
\renewcommand{\arraystretch}{1.9}
\centering
\footnotesize
\setlength\belowcaptionskip{5pt}
\caption{Wilcoxon test on the optimization DMM in relation to GEA vs Kingsley (time) and LEA (memory).} 
\label{Table:dmm_statistics} 
\begin{tabular}{lrrr} \hline
DMMs&\multicolumn{2}{c}{\textit{p-value}}\\ \hline 
GEA vs Kingsley (time) &\multicolumn{2}{r}{0.3951}\\
GEA vs LEA (memory)&\multicolumn{2}{r}{0.1415}\\ \hline
\end{tabular}
\end{table*}

Additionally, the DMM obtained with our GEA methodology was compared to Kingsley and LEA in memory use and execution time, respectively, in order to evaluate the relevance of our results regarding the measure that is not best for the reference DMMs. Table ~\ref{Table:dmm_statistics-2} shows results obtained. According to these tests, the memory use of GEA is better than Kingsley, and the Wilcoxon’s test demonstrates the relevance of the results obtained with a \textit{p-value} lower than $0.0018$. The comparison between LEA and GEA in terms of execution time also confirms that the results obtained by GEA are significant, with a \textit{p-value} lower than $0.002$), and it improves the performance with respect to LEA.

\begin{table*}[!t]
\renewcommand{\arraystretch}{1.9}
\centering
\footnotesize
\setlength\belowcaptionskip{5pt}
\caption{Wilcoxon test on the optimization DMM in relation to GEA vs Kingsley (memory) and LEA (time).} 
\label{Table:dmm_statistics-2} 
\begin{tabular}{lrrr} \hline
DMMs&\multicolumn{2}{c}{\textit{p-value}}\\ \hline 
GEA vs Kingsley (memory) &\multicolumn{2}{r}{0.001753}\\
GEA vs LEA (time)&\multicolumn{2}{r}{0.001988}\\ \hline
\end{tabular}
\end{table*}

\section{Conclusions and future work}\label{sec:Conclusions}
We have presented a method to optimize the memory subsystem of a computer addressing three different levels: register file, cache memory and dynamic memory management in the main memory. In all these levels we propose an evolutionary algorithm as the optimization engine, which is helped by other applications, either in a closed loop, either in off-line phases.

The optimization of the register file is based on a first step where a static profiling of the target applications is performed. Then, a multi-objective evolutionary algorithm is run, returning a set of solutions corresponding to register re-assignments. As a result, highly accessed registers are spaced far apart. In spite of thermal impact is not significant, we found some values worth to be studied and apply the optimization process. Our results obtain a reduction in the maximum temperature of $7.75\%$ and $10.79\%$ for some applications in ARM and VLIW architectures, respectively. This approach, as a consequence of reducing temperature, facilitates heat dissipation.

In the optimization of the single level cache memory we consider both the instructions and data caches, trying to reduce the execution time and the energy consumption due to cache memory operations. In this case, we propose a framework divided into three phases: two off-line phases responsible of cache characterization and the applications profiling, and a third phase which is driven by the evolutionary algorithm. The experiments return a set of cache configurations which, in terms of average execution time and average energy consumption, improve more than $34\%$ and $79\%$ respectively, compared with three baseline configurations. 

Regarding the dynamic memory management, our approach is divided into three phases: applications profiling, grammar generation and the optimization process based on Grammatical Evolution (GE). On average, we have obtained custom DMMs that improve the weighted function of execution time and memory use in a $59.27\%$, normalized with respect to the best general purpose DMMs.

The execution time of the experiments in the three memory layers was very diverse because it depends on the size of the target application, the configuration of the algorithm and the time devoted to evaluation of the different simulators that we call. Hence, our current optimization times are slow, and they are situated in the range of several hours for each experiment. Therefore, we will try to improve the execution time by incorporating parallel execution, as well as to fine-tune the configuration of the algorithms.

The proposed methodologies provide a useful framework to system designers facing the task of optimizing the memory subsystem of a device according to running applications. As future work we propose the integration of the three frameworks into a complete tool able to automatically optimize the three levels of the memory hierarchy without human interaction. In addition, we will study the use of artificial data in order to better understand the behavior of our algorithms, which could provide us clues to improve the tuning of our methods.


\bibliographystyle{ACM-Reference-Format}
\bibliography{tesis}
 
\end{document}